\documentclass[a4paper,12pt]{article}

\makeatletter
  
  \@addtoreset{equation}{section}
\makeatother

\begin{document}

\begin{titlepage}

\begin{flushright}
OU-HET 337\\
hep-th/9912096\\
December 1999
\end{flushright}
\vspace*{1.5cm}
\begin{center}
{\Large {\bf Three-Dimensional Extremal Black Holes and \\ the Maldacena
Duality}} \\
\bigskip
Naoto Yokoi\footnote{yokoi@het.phys.sci.osaka-u.ac.jp} and 
Toshio Nakatsu\footnote{nakatsu@het.phys.sci.osaka-u.ac.jp}\\
\bigskip
{\small
Department of Physics,\\
Graduate School of Science, Osaka University,\\
Toyonaka, Osaka 560, JAPAN
}
\end{center}
\bigskip
\bigskip
\bigskip

\begin{abstract}
We discuss the microscopic states of the extremal BTZ black holes. 
Degeneracy of the primary states
corresponding to the extremal BTZ black holes in the boundary N=(4,4)
SCFT is obtained by utilizing the elliptic genus and the unitary
representation theory of N=4 SCA.
The degeneracy is consistent with the Bekenstein-Hawking entropy.
\end{abstract}

\end{titlepage} 
 
\section{Introduction}

One of the challenging problems in quantum gravity is to understand the
microscopic properties of black holes, in particular, the statistical
origin of the Bekenstein-Hawking entropy.
New idea for an explanation of the 
origin of the Bekenstein-Hawking entropy has been provided by recent
development in our understanding of non-perturbative superstring theory.
It is based on the D-brane description of black
holes\cite{Strominger-Vafa} and the AdS/CFT 
correspondence\cite{Maldacena1,Witten,GKP}.
These are much related with each other under
the Maldacena duality\cite{Maldacena1}.

Three-dimensional Einstein equation with negative
cosmological constant has the solutions called 
the BTZ black holes\cite{BTZ1,BTZ2}.
These black holes have locally ${\rm AdS}_{3}$ geometry.
Via the ${\rm AdS}_{3}/{\rm CFT_{2}}$ correspondence, one can also 
hope to be able to analyze the microscopic properties of 
the BTZ black holes based on a local field theory on
the boundary. Infinite dimensional algebra of two-dimensional conformal
symmetry, that is, the Virasoro algebra, provides an important clue
for our understanding the correspondence and the Maldacena duality.
The pioneering work is Strominger's counting of microscopic states
of the BTZ black holes\cite{Strominger}.
But the qualitative aspects of this counting still remain obscure .

In this paper, we will discuss the three-dimensional extremal BTZ 
black holes in the context of the Maldacena duality.
Although this duality has been conjectural yet, various checks have been
carried out. (See \cite{Maldacena2} and references therein.)
In this perspective, the extremal BTZ black holes can be identified
with the primary states which are 1/2 BPS states in the N=(4,4)
two-dimensional supersymmetric $\sigma$-model.
This $\sigma$-model has a quantity called elliptic genus 
convenient to count the degeneracy of these states.
We explicitly count the microscopic states of the extremal BTZ black
holes with this identification by using the elliptic genus and
the unitary representation theory of the N=4 superconformal
algebra. The microscopic entropy of these black holes obtained by
this counting agrees with the entropy \'a la Bekenstein-Hawking.

This paper is organized as follows.
In section 2, we will summarize the
previous results about the BTZ black holes from the perspective of the
${\rm AdS}_{3}/{\rm CFT}_{2}$ correspondence in a pure quantum gravity and
in non-perturbative superstring theory, i.e., the Maldacena duality.
In section 3, after a brief introduction of N=4 superconformal algebra,
black hole states are discussed in the unitary representation
theory.
In section 4, some facts about the elliptic
genus of the N=(4,4) supersymmetric $\sigma$-model are reviewed.
In section 5, we count the number of 1/8 BPS states in the D1-D5 brane
system in IIB supergravity via the elliptic genus of this
$\sigma$-model and then finally count the microscopic states of the
extremal BTZ black holes.
In section 6, some other related topics are discussed.

\section{BTZ black holes and ${\bf AdS_{3}/CFT_{2}}$
correspondence}

\subsection{BTZ black holes in a three-dimensional pure quantum 
gravity}

The BTZ black holes
\footnote{Exact solutions of the vacuum 
Einstein equation with a negative cosmological 
constant $\Lambda=-1/l^2$.}
are three-dimensional black holes specified 
by their mass $M$ and angular momenta $J$, where $|J| \leq Ml$.   
In terms of the Schwarzschild coordinates 
$(t,\phi,r)$, with the ranges 
$-\infty<t<+\infty$, 
$0\leq \phi <2\pi$ 
and $0<r<+\infty$, 
the black hole metric $ds_{{\rm BTZ}_{(J,M)}}^2$ 
has the form 
\begin{eqnarray}
ds^{2}_{{\rm BTZ}_{(J,M)}} 
\equiv 
-N^2(dt)^2+N^{-2}(dr)^2
+r^2(d\phi+N^{\phi}dt)^2, 
\label{dsX(J,M)}
\end{eqnarray}
where $N$ and $N^{\phi}$ are the functions of the radial coordinate $r$ 
\begin{eqnarray}
N^2 =
\frac{(r^2-r_+^2)(r^2-r_-^2)}{l^2r^2},~~~~
N^{\phi} =  
\left\{ 
\begin{array}{cc}
\frac{r_+r_-}{lr^2} & \mbox{when}~~J \geq 0, 
\\ 
-\frac{r_+r_-}{lr^2} & \mbox{when}~~J < 0.  
\end{array}
\right. 
\label{N-Nphi}
\end{eqnarray}
The outer and inner horizons are located respectively 
at $r=r_+$ and $r=r_-$. Information about the mass 
and angular momentum is encoded in $r_{\pm}$ by  
\begin{eqnarray} 
r_{\pm}^2 \equiv 
4GMl^2 \left( 
1\pm \sqrt{1-\frac{J^2}{M^2l^2}} 
\right), 
\label{rpm}
\end{eqnarray}
where $G$ is Newton constant. 
In the case of $Ml=|J|$,  the black hole is called extremal. 
It holds $r_+=r_-$  
and then the outer and inner horizons coincide  
with each other. 
When $Ml>|J|$, it is called non-extremal.
And in the case of $J = 0$ and $Ml = -l/8 G$,
the geometry corresponds to the global ${\rm AdS}_{3}$.

The outer horizon of these solutions has finite area.
The semiclassical argument leads to
the finite Bekenstein-Hawking entropy:
\begin{eqnarray}
S \equiv \frac{A}{4 G} = \frac{2 \pi r_{+}}{4G}. \qquad (A : {\rm area
\ of \ the \ outer \ horizon})  \label{eq:entropy}
\end{eqnarray}

Quantization of three-dimensional
pure gravity with negative cosmological constant is discussed in 
\cite{Nakatsu}.
It is prescribed, through the detailed analysis of 
Brown-Henneaux's asymptotic Virasoro symmetry\cite{Brown-Henneaux}, 
as the geometric quantization of the Virasoro coadjoint orbits
of the Virasoro central charge $c=3 l/2 G$.

The BTZ black holes and the ${\rm AdS}_{3}$ correspond to the primary
states (highest weight states) 
of the Virasoro algebra of 
Brown-Henneaux\footnote{Strictly speaking,
these states correspond to the geometry of the exterior of outer
horizon of the BTZ black holes and the geometry without the origin of 
the ${\rm AdS_{3}}$ respectively}:
\begin{eqnarray}
{\rm BTZ}_{(J, M)} &\Longleftrightarrow& |J, M\rangle \equiv 
|h\rangle \otimes |\tilde{h}\rangle,  \label{eq:pribtz}\\  
{\rm AdS}_{3} &\Longleftrightarrow& |vac\rangle
\equiv |0\rangle\otimes|0\rangle, 
\label{eq:priads}
\end{eqnarray}
where
\begin{eqnarray}
h = \frac{1}{16Gl}(r_{+}+r_{-})^2 + \frac{c}{24}, \quad {\tilde h} = 
\frac{1}{16Gl}(r_{+}-r_{-})^2 + \frac{c}{24}.
\end{eqnarray}
The extremal BTZ black holes correspond to
\begin{eqnarray}
{\rm BTZ}_{(Ml, M)} \Longleftrightarrow |Ml, M\rangle \equiv 
|h \rangle \otimes |\frac{c}{24} \rangle. \label{eq:priext}
\end{eqnarray}
 
The total Hilbert space of the theory, which includes 
excited states (secondary states), 
is obtained by the tensor products ${\cal V}_{h}\otimes{\tilde {\cal
V}_{{\tilde h}}}$ of the Verma modules of the Virasoro algebra.
(${\cal V}_{h}$ and ${\tilde {\cal V}}_{{\tilde h}}$ are respectively
the Verma modules of the left-moving and right-moving sectors.) 
These Verma modules constitute the unitary irreducible 
representations of the Virasoro algebra. 
We can identify the states excited by $L_{-n}$ in the Verma module
with massive gravitons on the corresponding background geometry.

In view of the ${\rm AdS}_{3}/{\rm CFT}_{2}$ correspondence,
this Hilbert space should be realized by the corresponding boundary CFT. 
In fact it was done\cite{Nakatsu} based on 
the Liouville field $X$ with a specific
background charge. The action is given by
\begin{eqnarray}
S[X]=\frac1{4\pi i}\int_{{\bf P}^1} 
\bar{\partial} X \wedge \partial X
     + \frac{\alpha_0}{2\pi}\int_{{\bf P}^1} RX, 
\qquad \left( \alpha_0 \equiv \sqrt{\frac{l}{8G}} \right)
\end{eqnarray}
where $R$ is the Riemann tensor of 
a fixed K\"ahler metric on ${\bf P}^1$.
The stress tensor $T(z)$ has the form
\begin{eqnarray}
T(z)=-\frac{1}{2} 
\partial X \partial X(z)+\alpha_0\partial^2 X(z), 
\end{eqnarray}
and provides the generators 
of the Virasoro algebra  
with the central charge $1+12\alpha_0^2 = 1+3l/2G$.
This central charge is the same as that of Virasoro algebra of 
Brown-Henneaux in the semiclassical limit, i.e.,  $l/G \gg 1$.
The Fock space ${\cal F}_{k}$\footnote{Similar
arguments hold for the anti-holomorphic (right-moving) part.} is built on 
the Fock vacuum $|k\rangle$, which is introduced as 
the state obtained from the ordinary $SL_{2}({\bf C})$-invariant vacuum 
$|0\rangle$ by the relation
$|k\rangle = \lim_{z\rightarrow 0} e^{i k X(z)}|0\rangle$.

The BTZ black hole states (\ref{eq:pribtz}) can be 
identified with the following Fock vacuum:
\begin{eqnarray}
{\rm BTZ}_{(J, M)} \Longleftrightarrow |J, M\rangle \equiv 
  |k_{(J, M)}\rangle \otimes |\tilde{k}_{(J, M)}\rangle, 
\label{black hole state in 2d}
\end{eqnarray}
where $k_{(J,M)}$ and $\tilde{k}_{(J,M)}$ are given by 
\begin{eqnarray}
k_{(J, M)} &\equiv& 
-i\sqrt{\frac l{8G}}+\frac{r_+ +r_-}{\sqrt{8Gl}},
\nonumber \\ 
\tilde{k}_{(J, M)} &\equiv& 
 -i\sqrt{\frac l{8G}}+\frac{r_+ -r_-}{\sqrt{8Gl}}.   
\label{k(J,M)}
\end{eqnarray}
${\rm AdS}_{3}$ state (\ref{eq:priads}) can be  
identified with the $SL_{2}({\bf C})$-invariant vacuum:
\begin{eqnarray}
{\rm AdS}_{3} \Longleftrightarrow |vac\rangle
\equiv |0\rangle\otimes|0\rangle.
\end{eqnarray}

The Fock spaces ${\cal F}_{k} \otimes {\tilde {\cal
F}}_{{\tilde k}}$ built on these primary
states give the unitary irreducible representations of the Virasoro
algebra with $c=1+3l/2G$, and coincide with the physical Hilbert
space of the previous quantization of three-dimensional pure
gravity.

To summarize, in this correspondence of three-dimensional pure gravity
and the boundary CFT, the BTZ black holes appear as the primary
states of the Virasoro algebra with $c = 3l/2G$ in both descriptions.
This result may not be desirable for the counting of microscopic states of
the BTZ black holes. We cannot count in principle the
degeneracy of these primary states with this boundary theory,
since this Liouville field theory has continuum
spectrum of primary states.

\subsection{BTZ black holes and Maldacena duality}

Next, we consider the ${\rm AdS}_{3}/{\rm CFT}_{2}$ correspondence 
in superstring theory.
Through the analysis
of the near horizon limit of the BPS solitonic solution of 
IIB supergravity, which describes
the bound state of $Q_{1}$ D1-branes and $Q_{5}$ D5-branes,
Maldacena has conjectured in \cite{Maldacena1},
\begin{center}
IIB superstring theory on (${\rm AdS}_{3} \times {\rm S}^{3})_{Q_{1}Q_{5}} 
\times M_{4}$ \quad ($M_{4}$ = $K3$ or $T^4$)\\ 
$\Updownarrow$ dual  \\ two-dimensional N=(4,4) supersymmetric
$\sigma$-model \\ on the Higgs branch of world volume
theory of the D1-D5 system.
\end{center}
Here, we indicated the dependence of the radius of 
${\rm AdS}_{3}$ and ${\rm S}^{3}$ on $Q_{1}Q_{5}$ 
(see below). We call this duality simply the Maldacena duality.

We will discuss mainly the case of $M_{4}$ = $K3$ in the following.
The N=(4,4) $\sigma$-model can be regarded as
the $\sigma$-model on the target space of the $k$-th symmetric product of
$K3$ \cite{Vafa1,Dijkgraaf}, where
\begin{eqnarray}
k = Q_{1}Q_{5} + 1.
\end{eqnarray} 
Since the symmetric product 
is 4$k$-dimensional hyper-K\"ahler manifold, it
has automatically the N=(4,4) superconformal symmetry.
The Virasoro subalgebra and zero mode of the
SU(2) current algebra may
be identified with the Virasoro algebra of 
Brown-Henneaux on the boundary of ${\rm AdS}_{3}$ and 
the isometry of ${\rm S}^{3}$, respectively. We will discuss 
the N=4 superconformal algebra in more detail in section 3.

The extremal BTZ black holes can also be obtained
as the near horizon limit of the similar BPS solitonic solutions of 
IIB supergravity.
Therefore we can expect that the extremal BTZ black holes 
can be analyzed by this N=(4,4) $\sigma$-model.

We summarize some related facts of IIB supergravity here.
IIB supergravity on $S^{1} \times K3$ whose radius and volume are $R$
and $(2 \pi)^4 \alpha^{'2} v$ has the BPS solitonic solution
(see for example \cite{Skenderis} and references therein):
\begin{eqnarray}
ds_{10}^{2} &=& f_{1}^{-\frac{1}{2}}f_{5}^{-\frac{1}{2}}\{ - dt^2 +
dx_{5}^{2} + f_{N}(dt+dx_{5})^2\} \nonumber \\ && + f_{1}^{\frac{1}{2}}
f_{5}^{\frac{1}{2}}(dx_{1}^{2}+dx_{2}^{2}+dx_{3}^{2}+dx_{4}^{2}) + 
f_{1}^{\frac{1}{2}}f_{5}^{-\frac{1}{2}} ds_{K3}^{2}, \label{eq:10dim} 
\end{eqnarray}
with periodic identification $x_{5} \sim x_{5} + 2 \pi R$
in string frame and
\begin{eqnarray}
e^{-2 (\phi - \phi_{\infty})} &=& f_{5}f_{1}^{-1}, 
\quad C_{05}^{(R)} = \frac{1}{2}(f_{1}^{-1} - 1), \nonumber \\
H_{ijk}^{(R)} &=& (*_{6} \ dC^{(R)})_{ijk} =
\frac{1}{2}\epsilon_{ijkl}\partial_{l}f_{5} \qquad (i,j,k,l =
1,2,3,4),
\end{eqnarray}
where $C^{R}$ is Ramond-Ramond 2-form and $*_{6}$ is Hodge dual in
6-dimension ($t,x_{1},\cdots,x_{5}$).
And $f_{1}$, $f_{5}$ and $f_{N}$ are following functions with respect
to radial coordinate, \\
$r = x_{1}^{2}+x_{2}^{2}+x_{3}^{2}+x_{4}^{2}$:
\begin{eqnarray}
f_{1} &=& 1 + \frac{{\tilde Q}_{1}}{r^2}, \qquad {\tilde Q}_{1} =
\frac{\alpha^{'} g_{st}}{v} Q_{1}, \\
f_{5} &=& 1 + \frac{{\tilde Q}_{5}}{r^2}, \qquad {\tilde Q}_{5} = 
\alpha^{'} g_{st}Q_{5}, \\
f_{N} &=& \frac{{\tilde N}}{r^2}, \qquad \qquad {\tilde N} =
\frac{\alpha^{'2} g_{st}^{2}}{R^{2} v} N.
\end{eqnarray}

This solution corresponds to the configuration
of the bound state of $Q_{5}$ D5-branes wrapping on $K3 \times S^{1}$
and $Q_{1}$ D1-branes wrapping on $S^{1}$ with $N$ units of KK momenta
along $S^{1}$, i.e., $x_{5}$-direction, and preserves four
supercharges. So it is a 1/8 BPS state\cite{Strominger-Vafa}.

Here we can get the extremal BTZ black hole as the near horizon limit of
the geometry (\ref{eq:10dim})\cite{Maldacena-Strominger,Skenderis}.  
The near horizon limit is defined as
\begin{eqnarray*}
\alpha{'} \rightarrow 0, \qquad {\rm with} \qquad  
U \equiv \frac{r}{\alpha^{'}}, \  R  \  {\rm and}  \   v  \
  {\rm fixed}.
\end{eqnarray*}
In this limit, the metric (\ref{eq:10dim})
describes $({\rm BTZ}_{(Ml,M)} \times S^{3})_{Q_{1}Q_{5}}
\times K3$ with $Ml = J = N$.
The radius of ${\rm AdS}_{3}$ and ${\rm S}^{3}$ coincide and become
$l = g_{st}^{1/2} \alpha^{' 1/2} (Q_{1}Q_{5}/v)^{1/4}$. \\
And the three-dimensional effective Newton
constant on ${\rm BTZ}_{(N,N/l)}$ is given by
$G_{{\rm eff}}^{(3)} = l/(4 Q_{1} Q_{5})$.
There exists on
this background the asymptotic Brown-Henneaux's Virasoro symmetry
with central charge,
\begin{eqnarray}
c = \frac{3l}{2G_{{\rm eff}}^{(3)}} = 6 Q_{1}Q_{5}.
\end{eqnarray}
This is the same as the central charge of the N=(4,4) $\sigma$-model 
at the semiclassical limit $Q_{1}Q_{5} \gg 1$. 
The Bekenstein-Hawking entropy becomes
\begin{eqnarray}
S = \frac{2 \pi r_{+}}{4 G_{{\rm eff}}^{3}} = 2 \pi
\sqrt{Q_{1}Q_{5}N}, \label{eq:extentropy}
\end{eqnarray}
which is valid in the semiclassical region $N \gg Q_{1}Q_{5} \gg 1$.

If one accepts the Maldacena duality, the extremal black hole should be
identified with the primary state
\begin{eqnarray}
|N+\frac{c}{24}\rangle \otimes |\frac{c}{24} \rangle, \label{eq:extprimary}
\end{eqnarray}
of the N=(4,4) $\sigma$-model.
One can ask whether the entropy (\ref{eq:extentropy}) can be regarded as 
the degeneracy of the primary state (\ref{eq:extprimary}) of this N=(4,4)
$\sigma$-model. In the sequel, we will discuss this question and
answer in the affirmative. 

\section{N=4 superconformal symmetry}
N=(4,4) $\sigma$-model is known to be finite to all orders of
perturbation and to be conformally invariant at the quantum
level. Thus the states of the $\sigma$-model on the $k$-th symmetric
product\footnote{We will
denote $k$-th symmetric product of $K3$ as $S^k K3 \equiv
K3^{\otimes k}/S_{k} \ (S_{k}$ is a $k$-dimensional symmetric
group).}, $S^k K3$, constitute the unitary
irreducible representations of the underlying N=4 superconformal
algebra (N=4 SCA).

\subsection{Basics of N=4 superconformal algebra}
N=4 SCA is generated by $L_{n},\ J_{n},\ G_{r}^{i}$ and ${\bar
G}_{r}^{i}$ with 
\begin{eqnarray*}
[L_{m},L_{n}] &=& (m-n) L_{m+n}+\frac{k}{2}m(m^2-1)\delta_{n+m,0}, \quad
\{G_{r}^{i},G_{s}^{j}\} = \{\bar{G}_{r}^{i},\bar{G}_{s}^{j}\} = 0, \\
\{G_{r}^{i},{\bar G}_{s}^{j}\} &=& 2
\delta^{ij}L_{r+s}-2(r-s)\sigma_{ij}^{a}J_{r+s}^{a} + \frac{k}{2}(4
r^2-1) \delta_{r+s,0}, \\
\left[ J_{m}^{a},J_{n}^{b} \right] &=& i \epsilon^{abc} J_{m+n}^{c} + 
\frac{k}{2}m\delta_{m+n,0}, \\
\left[ J_{m}^{a},G_{r}^{i} \right] &=&
-\frac{1}{2}\sigma_{ij}^{a}G_{m+r}^{j}, \quad 
\left[ J_{m}^{a},{\bar G}_{r}^{i} \right] =
\frac{1}{2}\left(\sigma_{ij}^{a}\right)^{*}{\bar G}_{m+r}^{j}, \\
\left[ L_{m},G_{r}^{i} \right] &=&
(\frac{m}{2}-r)G_{m+r}^{i}, \quad \left[ L_{m},{\bar G}_{r}^{i} \right] =
(\frac{m}{2}-r){\bar G}_{m+r}^{i}, \\
\left[ L_{m},J_{n}^{a} \right] &=& -n J_{m+n}^{a}. 
\end{eqnarray*}
($\sigma_{ij}^{a}$ is Pauli matrix and $J_{n}^{(\pm)} \equiv 
J_{n}^{1}\pm i J_{n}^{2}$.) $L_{n},J_{n}^{a}$ and 
$G_{r}^{i}({\bar G}_{r}^{i})$ represent the
Fourier components of the energy momentum tensor, SU(2) current and four
supercurrents, respectively. 
$G_{r}^{i} ({\bar G}_{r}^{i})$ transforms as SU(2) doublet (its conjugate) 
under the global SU(2) symmetry which is generated by $J_{0}^{a}$.
The level $k$ of the SU(2) current algebra ($\widehat{{\rm SU}(2)}_{k}$)
must be positive integer for 
unitary representations. The central charge $c$ of the Virasoro subalgebra is 
$6k$. 

Two different boundary conditions of the supercurrents
provide two different sectors of this algebra. 
If $G_{r}^{i}({\bar G}_{r}^{i})$ has $r \in
{\bf Z}+ \frac{1}{2} $, it is called the Neveu-Scwarz (NS) sector 
and if $r \in {\bf Z}$, it is called the Ramond (R) sector.
These sectors are related by the automorphism of the algebra
called spectral flow.
We will discuss the R-sector in the following.

Unitary irreducible representations of N=4 SCA have
two distinct types\cite{Eguchi1}.
They are built on highest weight states $|h,l\rangle$ called massive
primary and massless primary in the R-sector.
\begin{enumerate}
\renewcommand{\labelenumi}{\theenumi}
\renewcommand{\theenumi}{(\roman{enumi})}
\item Massive primary state
      \begin{eqnarray}
      L_{n}|h,l\rangle &=& G_{n}^{i}|h,l\rangle = 
      {\bar G}_{n}^{i}|h,l\rangle = J_{n}^{a}|h,l\rangle = 0, \qquad n
      \geq 1 \nonumber \\
      J_{0}^{(+)}|h,l\rangle &=& G_{0}^{2} |h,l\rangle = 
      {\bar G}_{0}^{1}|h,l\rangle = 0, \nonumber \\
      L_{0}|h,l\rangle &=& h |h,l\rangle, \quad J_{0}^{3} |h,l\rangle =
      l |h,l\rangle, \nonumber \\
      h &>& \frac{k}{4} = \frac{c}{24}, 
      \qquad l = \frac{1}{2},1,\cdots,\frac{k}{2}-\frac{1}{2},\frac{k}{2}. 
      \label{eq:massive}
      \end{eqnarray}
\item Massless primary state
      \begin{eqnarray}
      L_{n}|h,l\rangle &=& G_{n}^{i}|h,l\rangle = 
      {\bar G}_{n}^{i}|h,l\rangle = J_{n}^{a}|h,l\rangle = 0, \qquad n
      \geq 1 \nonumber \\
      J_{0}^{(+)}|h,l\rangle &=& G_{0}^{i} |h,l\rangle = 
      {\bar G}_{0}^{i}|h,l\rangle = 0, \qquad i=1,2 \nonumber \\
      L_{0}|h,l\rangle &=& h |h,l\rangle, \quad J_{0}^{3} |h,l\rangle =
      l |h,l\rangle, \nonumber \\
      h&=&\frac{k}{4}=\frac{c}{24},\qquad l =0,\frac{1}{2},\cdots,
      \frac{k}{2}-\frac{1}{2},\frac{k}{2}. \label{eq:massless} 
      \end{eqnarray}
\end{enumerate}
These representations are called massive representation ${\cal M}^{k}_{(h,l)}$ 
and massless representation ${\cal M}^{k}_{0 (l)}$ respectively.

The massive representations have the same number of bosonic and
fermionic states at each level and the Witten index is equal to 
zero. 
These representations correspond to
the representations which have spontaneously broken supersymmetry.
The Witten index of the massless representations is non-zero. 
These representations are the representations which have
unbroken supersymmetry. 
The primary states of massless representations have dimension 
$h=k/4=c/24$. These are the ground states of the R-sector.

Character of the representation is introduced by
${\rm ch}^{(R)}(\tau,z) = 
{\rm Tr} (q^{L_{0}-\frac{c}{24}} y^{2 J_{0}^{3}})$.
($q=e^{2 \pi i \tau}$ and $y=e^{2 \pi i z}$.)  
Their explicit form is given in \cite{Eguchi2}.     
\begin{enumerate}
\renewcommand{\labelenumi}{\theenumi}
\renewcommand{\theenumi}{(\roman{enumi})}
\item The character of the massive representation ${\cal M}^{k}_{(h,l)}$: 
      \begin{eqnarray}
      {\rm ch}^{(R) k}(h,l ; \tau,z) =
      q^{h-\frac{k}{4}-\frac{l^2}{k+1}}
      \frac{\theta_{2}(\tau,z)^2}{\eta(\tau)^3}
      \chi_{k-1}^{l-\frac{1}{2}}(\tau,z),  \label{eq:charmass}
      \end{eqnarray}
      where $\chi_{k}^{l}(\tau,z)$ is the character of 
      $\widehat{{\rm SU}(2)}_{k}$ of isospin $l$,
      \begin{eqnarray}
      \chi_{k}^{l}(\tau,z) &=& \frac{q^{\frac{(l+\frac{1}{2})^2}{k+2}
      -\frac{1}{8}}}{\prod_{n=1}^{\infty}(1-q^n)(1-y^2
      q^n)(1-y^{-2}q^{n-1})} \nonumber \\
      &&\times \sum_{m=0}^{\infty}q^{(k+2)m^2 + (2 l + 1)}\left(y^{2
      \{(k+2)m + l\}}-y^{-2 \{ (k+2)m+l+1 \}}\right) \nonumber \\
      && = \frac{\Theta_{2l+1, k+2}(\tau, 2z) - \Theta_{-2l-1, k+2}(\tau,
        2z)}{\Theta_{1,2}(\tau,2z) - \Theta_{-1,2}(\tau,2z)}.
      \label{eq:charaffine}
      \end{eqnarray}
      $\Theta_{l,k}(\tau,z) = \sum_{n=-\infty}^{\infty} q^{k
      (n+\frac{l}{2k})^2} y^{k (n+\frac{l}{2k})}$ is the theta
      function associated with $\widehat{{\rm SU}(2)}_{k}$ of isospin $l$.
\item The character of the massless representation ${\cal M}^{k}_{0 (l)}$:
      \begin{eqnarray}
      {\rm ch}_{0}^{(R) k}(h=\frac{k}{4},l;\tau,z) &=&
      q^{-\frac{1}{8}}\frac{\theta_{2}(\tau,z)^2}{\eta(\tau)^3} 
      \frac{1}{\prod_{n=1}^{\infty}(1-q^n)(1-y^2
      q^n)(1-y^{-2}q^{n-1})} \nonumber \\
      \times \sum_{m=-\infty}^{\infty}&& \hskip-1cm q^{(k+1)m^2 + 2 l m} 
      \left( \frac{y^{2 \{(k+2)m+l-\frac{1}{2}\}}}{(1+y^{-1}q^{-m})^2} - 
      \frac{y^{-2 \{ (k+2)m+l+\frac{1}{2}\}}}{(1+y q^{-m})^2} \right).
      \label{eq:charless}
      \end{eqnarray}
\end{enumerate}

These characters enjoy the following properties.
The Witten index of the representation 
can be obtained, if one sets $z=1/2$, i.e., $y=-1$:
\begin{eqnarray}
{\rm ch}^{(R) k}(h,l;\tau,z=\frac{1}{2}) &=& 0, \\
{\rm ch}_{0}^{(R) k}(h=\frac{k}{4},l;\tau,z=\frac{1}{2}) &=& 
(-1)^{2 l}(2 l+1).
\end{eqnarray}
The characters of massive and massless
representations are related by
\begin{eqnarray}
{\rm ch}^{(R) k}(h=\frac{k}{4},l;\tau,z) &=&  
{\rm ch}_{0}^{(R) k}(h=\frac{k}{4},l;\tau,z)  
+ 2 \ {\rm ch}_{0}^{(R) k}(h=\frac{k}{4},l-\frac{1}{2};\tau,z)
\nonumber \\
&&+ \ {\rm ch}_{0}^{(R) k}(h=\frac{k}{4},l-1;\tau,z).
\label{eq:relation}
\end{eqnarray}

\subsection{Identification of the black hole states}

As argued in section 2.2, the extremal BTZ black hole will correspond
to the primary state (\ref{eq:extprimary}) of the N=(4,4)
$\sigma$-model.
It is a Virasoro primary state of the underlying N=4
SCA. The fact that the extremal BTZ black holes are the 1/2 
BPS states with respect to the Poincare supersymmetry in 
three dimensions\footnote{These correspond to the 1/4
BPS states in Anti-de Sitter supersymmetry in 3-dimension.} implies
that this primary state is in the tensor
product of massive and massless representations ${\cal
M}^{k}_{(h,l)} \otimes {\tilde {\cal M}}^{k}_{0 ({\tilde l})}$ which
is built on 
\begin{eqnarray}
|h=N+\frac{k}{4},l\rangle \otimes |{\tilde h}=\frac{k}{4},{\tilde
 l}\rangle. \label{eq:prisigma}
\end{eqnarray}

Actually we can proceed further. Since the extremal 
BTZ black holes do not have the conserved 
charge corresponding to isospin $l$ and ${\tilde l}$,
the primary state (\ref{eq:extprimary}) may be identified with the
Virasoro primary state having vanishing isospins $l={\tilde
l}=0$ in 
${\cal M}^{k}_{(h,l)} \otimes {\tilde {\cal M}}^{k}_{0 ({\tilde l})}$.
The primary states with $l=0$ 
in ${\cal M}^{k}_{(h,l)}$
are as follows:
when $l \in {\bf Z}$, they are given by 
\begin{eqnarray}
&&J_{0}^{(-) \ l} |h=N+\frac{k}{4},l\rangle, \nonumber \\
{\rm and} && J_{0}^{(-) \ l-1} ({\bar G}_{0}^{2} G_{0}^{1} - 
\frac{h-k/4}{l}) |h=N+\frac{k}{4},l \rangle, \label{eq:integer}
\end{eqnarray}
and when $l \in {\bf Z}+\frac{1}{2}$, they are 
\begin{eqnarray}
&&J_{0}^{(-) \ l-\frac{1}{2}} G_{0}^{1} |h=N+\frac{k}{4},l\rangle,
\nonumber \\
\hskip-1.4cm{\rm and} \qquad 
&&J_{0}^{(-) \ l-\frac{1}{2}}{\bar
G}_{0}^{2}|h=N+\frac{k}{4},l\rangle.
\label{eq:halfint} 
\end{eqnarray}
The primary states with ${\tilde l}=0$ in
${\tilde {\cal M}}^{k}_{0 ({\tilde l})}$ is identified with
${\bar J}_{0}^{(-) \ {\tilde l}} |{\tilde h}=k/4,{\tilde
l}\rangle.$
The primary state (\ref{eq:extprimary}) can be identified with the
tensor product of these states. So, the degeneracy of the state is
almost same as the degeneracy of the representation ${\cal M}^{k}_{(h,l)} 
\otimes {\tilde {\cal M}}^{k}_{0 ({\tilde l})}$. 
  
\section{Elliptic genus for $\sigma$-model on symmetric product
of K3}
To count the number of 1/2 BPS states in the N=(4,4)
$\sigma$-model,
the so-called ``elliptic genus'' is a convenient tool.
We summarize some properties of the elliptic genus
emphasizing its modular transform and examine it from the perspective
of N=4 SCA. 
\subsection{Elliptic genus as a weak Jacobi form}
The elliptic genus of target space $M$ 
is defined by the following trace in the R-R sector of the underlying
N=(2,2) superconformal field theory\footnote{One can obtain the following
topological indices of the target space $M$, if one sets $z$ to be
specific value.
\begin{eqnarray}
Z[M](\tau,0) &:& {\rm Elliptic \ extension \ of \ Euler \ number}
\nonumber \\
Z[M](\tau,\frac{1}{2})&:& {\rm Elliptic \ extension \ of \
Hirzebruch \ signature} \nonumber \\
q^{\frac{c}{24}}Z[M](\tau,\frac{\tau+1}{2})&:& 
{\rm Elliptic \ extension \ of \ Dirac \ genus} \nonumber
\end{eqnarray}}.
\begin{eqnarray}
Z[M](\tau,z) = {\rm Tr}_{{\rm R}\textrm{-}{\rm R}}(-1)^{J_{0}-{\bar
J}_{0}} q^{L_{0}-\frac{c}{24}} {\bar q}^{{\bar
L}_{0}-\frac{c}{24}} y^{J_{0}}, \label{eq:genus}
\end{eqnarray}
where $J_{0}$ and ${\bar J}_{0}$ are the integral N=2 U(1) charges
of the left-moving and the right-moving sectors\footnote{N=2 
SCA can be embedded into N=4 SCA by
$G_{r} = G_{r}^{1} + {\bar G}_{r}^{2}$, ${\bar G}_{r} = 
G_{r}^{2} + {\bar G}_{r}^{1}$ and $J_{n}=2 J_{n}^{3}$.}.   
The elliptic genus is
independent of ${\bar \tau}$ by virtue of supersymmetry of the R-sector.
The contribution of the right-moving sector is only from the
ground states.
But all states in the left-moving sector contribute to $Z[M](\tau,z)$.
So the elliptic genus $Z[M](\tau,z)$ is a useful quantity for 
counting of the 1/2 BPS states.

The following theorem is known about this elliptic genus. (See
 \cite{Kawai2} for detail.)

\noindent
{\bf Theorem 1.} If the target space $M$ of the $\sigma$-model is an
even-dimensional Calabi-Yau manifold, then the elliptic genus
$Z[M](\tau,z)$ is a weak Jacobi form of weight 0 and index
$d/2 (d=\dim_{{\bf C}} M)$ without character.

Weak Jacobi form in the above theorem is defined as
follows\cite{Eichler}.

\noindent 
{\bf Definition.} A function 
$\phi(\tau,z)$ is called a weak Jacobi form of weight $k \in {\bf Z}$ 
and index $m \in {\bf Z}_{>0}/2$ without character, if it satisfies
(i)$\sim$(iv): 
\begin{enumerate}
\renewcommand{\labelenumi}{\theenumi}
\renewcommand{\theenumi}{(\roman{enumi})}
\item $\phi(\tau,z)$ is a holomorphic function with respect to $\tau
      \in {\bf H}^{+} \ ({\bf H}^{+}:$ upper half plane) and $z \in
      {\bf C}$.
\item $\phi(\frac{a \tau + b}{c \tau + d},\frac{z}{c \tau + d}) = 
      (c \tau + d)^{k}e^{\frac{2 \pi i m c z^{2}}{c \tau +
      d}}\phi(\tau,z). \quad (a,b,c,d \in {\bf Z} \ {\rm and} \ ad-bc=1)$
\item $\phi(\tau,z + \lambda \tau + \mu) = e^{- 2\pi i m (\lambda^2
      \tau + 2 \lambda z)} \phi(\tau,z). \quad (\lambda, \mu \in {\bf Z})$
\item $\phi(\tau,z)$ has the Fourier expansion of the form \\
      \begin{eqnarray*}
      \phi(\tau,z) = \sum_{n=0}^{\infty} \sum_{r = - \infty}^{\infty}
      c(n,r) q^n y^r \quad (q= e^{2 \pi i \tau},y = e^{2 \pi i z}).
      \end{eqnarray*}
\end{enumerate}

When $M$ is $K3$, the elliptic genus $Z[K3](\tau,z)$
becomes a weak Jacobi form of weight 0 and index 1 without character.
An actual calculation of the N=(4,4) supersymmetric
$\sigma$-model on $K3$\cite{Eguchi4,Kawai1} determines it explicitly as
\begin{eqnarray}
Z[K3](\tau,z) = 24 \wp(\tau,z)K^2(\tau,z), \label{eq:pfnK3}
\end{eqnarray}
where
\begin{eqnarray}
\wp(\tau,z) &:& {\rm Weierstrass's} \ 
\wp {\rm -function}, \nonumber \\
K(\tau,z) &=& i \frac{\theta_{1}(\tau,z)^2}{\eta(\tau)^3}.
\label{eq:kfn} 
\end{eqnarray} 

We need the following theorem about weak Jacobi form\cite{Kawai2,Eichler}. 

\noindent
{\bf Theorem 2}. If we assign weights 4, 6 and 2 respectively to
$E_{4}(\tau), E_{6}(\tau)$\footnote{$E_{4}(\tau)$ and $E_{6}(\tau)$
are the Eisenstein series.} and $\wp(\tau,z)$, any weak Jacobi
form of weight $2l$ ($l \in {\bf Z}_{\geq 0}$) and index $k$ ($k \in {\bf
Z}_{\geq 0}$) can be expressed as 
\begin{eqnarray*}
{\cal G}_{2 l + 2 k}(E_{4}(\tau), E_{6}(\tau), \wp(\tau,z))
K^{2k}(\tau,z),
\end{eqnarray*}
where ${\cal G}_{2l + 2k}(E_{4},E_{6},\wp)$ is a homogenious 
polynomial of weight (2$l$ + 2$k$) and its degree as a
polynomial in $\wp$ is at most $k$.

$S^k K3$ is 2$k$-dimensional Calabi-Yau manifold. 
The elliptic genus
$Z[S^{k}K3](\tau,z)$ becomes a weak Jacobi form of weight 0 and index $k$.   
According to theorem 2, it has the
following form:
\begin{eqnarray}
Z[S^k K3](\tau,z) = {\cal G}_{2k}(E_{4}(\tau), E_{6}(\tau),
\wp(\tau,z))K^{2k}(\tau,z). \label{eq:generalgenus}
\end{eqnarray}  

The homogenious polynomial ${\cal G}_{2k}$ is determined for lower
values of $k$\cite{Kawai2}\footnote{It is worth commenting that 
the coefficient of the first term in the
bracket is the Euler number of $S^k K3$, $\chi(S^k K3)$.}.
\begin{eqnarray}
k=1 &:& 24\wp K^{2} \nonumber \\
k=2 &:& (324 \wp^{2} + \frac{3}{4} E_{4})K^{4} \nonumber \\
k=3 &:& (3200 \wp^{3} + \frac{64}{3}E_{4}\wp +
\frac{10}{27}E_{6})K^{6}
\label{eq:pfnprod}
\end{eqnarray}

\subsection{Elliptic genus of $S^k K3$ and characters of N=4
superconformal algebra}
The N=(4,4) $\sigma$-model on the target space of $K3$ has been
analyzed in detail by Eguchi et.al.\cite{Eguchi4} in the context of 
a compactification of string theory on $K3$. The elliptic extension
of the Hirzebruch signature of $K3$ 
can be represented by the characters of the N=4
SCA as
\begin{eqnarray}
&&\hspace{-2cm} Z[K3](\tau,\frac{1}{2}) \nonumber \\
&=&-2 \ {\rm ch}_{0}^{(R) k=1}(h=\frac{1}{4},l=\frac{1}{2};\tau,0) +
20 \ {\rm ch}_{0}^{(R) k=1}(h=\frac{1}{4},l=0;\tau,0) \nonumber \\
&&+ F(\tau) \ {\rm ch}^{(R) k=1}(h=\frac{1}{4},l=\frac{1}{2};\tau,0)
\nonumber \\
&=& 24 \ {\rm ch}_{0}^{(R) k=1}(h=\frac{1}{4},l=0;\tau,0) +
 {\tilde F}(\tau) \ {\rm ch}^{(R)
k=1}(h=\frac{1}{4},l=\frac{1}{2};\tau,0),
\label{eq:genusK3}
\end{eqnarray}
where
\begin{eqnarray}
{\tilde F}(\tau) = -2 + F(\tau) = \sum_{n=0}^{\infty}a_{n}q^{n} \qquad
(a_{0} = -2, \ a_{n} \in {\bf Z}_{\geq 0} (n>0)).
\end{eqnarray} 
We have used eq.(\ref{eq:relation}) to obtain the last equality in 
(\ref{eq:genusK3}).
The degeneracy of the massive primary states in the left-moving sector
is encoded in ${\tilde F}(\tau)$. The coefficient $a_{n}$ is the
degeneracy of the massive primary states of $h=n+1/4$.

The function ${\tilde F}(\tau)$ can be
determined by combining
two expressions of elliptic genus (\ref{eq:pfnK3}) and (\ref{eq:genusK3}),
\begin{eqnarray}
Z[K3](\tau,\frac{1}{2}) &=& 24 \ {\rm ch}_{0}^{(R) k=1}
(h=\frac{1}{4},l=0;\tau,0) \nonumber \\
&& + {\tilde F}(\tau) \ {\rm ch}^{(R)
k=1}(h=\frac{1}{4},l=\frac{1}{2};\tau,0) \nonumber \\
&=& 24 \ \wp(\tau,\frac{1}{2})K^{2}(\tau,\frac{1}{2}).
\end{eqnarray} 
This gives
\begin{eqnarray}
{\tilde F}(\tau) &=& 2 \ \frac{\theta_{2}(\tau,0)^4 -
\theta_{4}(\tau,0)^4}{\prod_{n=1}^{\infty}(1-q^n)^3} - 24 \  
{\tilde h}_{3}(\tau), 
\end{eqnarray}
where
\begin{eqnarray} 
{\tilde h}_{3}(\tau) &=&
\frac{1}{\theta_{3}(\tau,0)}\sum_{m=-\infty}^{\infty}
\frac{q^{\frac{m^2}{2}}}{1+q^{m-\frac{1}{2}}}.
\end{eqnarray}

Now, we will turn to the case of the symmetric product.
The elliptic genus of $S^k K3$ can be 
also expanded by the characters of the underlying N=4 SCA.
In general, it has the following expansion:
\begin{eqnarray}
&&\hspace{-2cm}Z[S^k K3](\tau, z+\frac{1}{2}) 
\qquad \left( = {\rm Tr}(-1)^{- 2 {\bar
J}_{0}^{3}}{\bar q}^{{\bar
L}_{0}-\frac{c}{24}}q^{L_{0}-\frac{c}{24}}y^{2 J_{0}^{3}} \right) \nonumber \\
&&\hspace{-1cm}= \chi(S^k K3) \ {\rm ch}_{0}^{(R) k}
(h=\frac{k}{4},l=0;\tau,z) + 
\sum_{l=\frac{1}{2}}^{\frac{k}{2}}F_{l}(\tau) \ {\rm ch}^{(R) k}
(h=\frac{k}{4},l;\tau,z), \label{eq:charfnprod}
\end{eqnarray}
where
\begin{eqnarray}
F_{l}(\tau) = \sum_{n=0}^{\infty}a_{n}^{(l)}q^n, \qquad \quad (a_{0}^{(l)}
\in {\bf Z}, \ a_{n}^{(l)} \in {\bf Z}_{\geq 0} \ (n>0)). \label{eq:ffun}
\end{eqnarray}        
Although the characters of the massless representations with $l \neq
0$ may appear in $Z[S^k K3](\tau,z+1/2)$, we can reduce them
to $l=0$ by applying (\ref{eq:relation}) recursively. $a_{n}^{(l)}$ in
(\ref{eq:ffun}) provides the degeneracy of the representation 
${\cal M}_{(h,l)}^{k} \otimes {\tilde {\cal M}}_{0 ({\tilde l})}^{k}$ 
with $h=n+k/4$ and isospin $l$. $\sum_{l}a_{n}^{(l)}$ provides 
the number of the representations ${\cal M}_{(h,l)}^{k} 
\otimes {\tilde {\cal M}}_{0 ({\tilde l})}^{k}$ with $h=n+k/4$. 

Because of eq.(\ref{eq:generalgenus}), 
the spectrum of the massive primary states of this $\sigma$-model on 
$S^k K3$ also
becomes discrete and all states have dimension $h = {\bf Z}_{\geq 0}
+k/4$ in $Z[S^k K3](\tau,z+1/2)$. This corresponds to the fact 
the extremal BTZ black holes have the discrete mass $N/l$ in the context of
the Maldacena duality. 

The function $F_{l}(\tau)$ \ (or $\sum_{l}F_{l}(\tau)$) in
eq.(\ref{eq:charfnprod}) can be determined in principle by an 
analogous way with the case of $K3$.
But it is a hard task to obtain the exact functional form of 
$Z[S^k K3](\tau,z)$ such as (\ref{eq:pfnprod}) for the case of 
general $k$. 
However, as discussed in section 2.2, what we need for 
the counting of the microscopic states comparable with 
the Bekenstein-Hawking entropy is   
the asymptotic form of $a_{n}^{(l)}$ (or $\sum_{l} a_{n}^{(l)}$) 
at $n \rightarrow \infty$, since eq.(\ref{eq:extentropy}) is valid
for the region $N \gg Q_{1}Q_{5} \gg 1$.
We will consider this asymptotic form in the next section.

\section{State counting via the N=(4,4) $\sigma$-model}   
In this section, we will discuss the degeneracy of  
1/8 BPS states of the D1-D5
system in IIB superstring theory and the degeneracy of the primary states 
corresponding to the extremal BTZ black holes.
In the previous section, we obtain two different expressions
of the elliptic genus of $S^k K3$. 
We will first use the expression in terms of a weak
Jacobi form and count the number of the 1/8 BPS states by using a
Tauberian theorem. Then using the expression in terms of the 
characters of N=4 SCA, we will discuss the
degeneracy of the massive primary states and obtain the
microscopic entropy of the corresponding extremal BTZ black holes.

\subsection{Counting the 1/8 BPS states}
Let us start by studying the asymptotic behavior of the elliptic genus
$Z[S^k K3](\tau,1/2)$ as $\tau \downarrow 0$\footnote{
$\tau \downarrow 0 \stackrel{{\rm def}}{\Longleftrightarrow} \tau = iT
\ (T \in {\bf R}_{>0}),\  {\rm and} \ T \rightarrow 0.$}. 
Due to the structure theorem the elliptic genus
has the form (\ref{eq:generalgenus})
\begin{eqnarray*}
Z[S^k K3](\tau,\frac{1}{2}) = \
{\cal G}_{2 k}\left( E_{4}(\tau),E_{6}(\tau),\wp(\tau,\frac{1}{2})
\right) K^{2 k}(\tau,\frac{1}{2}). 
\end{eqnarray*}
The asymptotics can be obtained 
from those of the constituents in (\ref{eq:generalgenus}).
$\wp(\tau,1/2)$, $E_{4}(\tau)$ and
$E_{6}(\tau)$ behave as 
$\wp(\tau,1/2) \rightarrow (-1/12)(-i \tau)^{-2}, \
E_{4}(\tau) \rightarrow 1 (-i \tau)^{-4}$, and   
$E_{6}(\tau) \rightarrow (-1)(-i \tau)^{-6}$. 
Therefore the asymptotics of ${\cal G}_{2k}$ becomes
\begin{eqnarray}
{\cal G}_{2k}(\tau,\frac{1}{2}) \rightarrow {\tilde c}(k) (-i \tau)^{-2 k}
\qquad {\rm as} \ \tau \downarrow 0,
\end{eqnarray}
where ${\tilde c}(k)$ is a constant which depends on $k$ and the
polynomial form of ${\cal G}_{2k}$.
The asymptotics of $K^{2}(\tau,1/2)$ can be read from 
$\theta_{2}(\tau,0) \rightarrow 1 (-i \tau)^{-\frac{1}{2}}$ and 
$\eta(\tau) \rightarrow 1 (-i \tau)^{-\frac{1}{2}} e^{-\frac{\pi i}{12
\tau}}$,
\begin{eqnarray}
K^{2}(\tau,\frac{1}{2}) \rightarrow (-1)(-i \tau)^{2} e^{\frac{\pi i}{2
\tau}} \qquad {\rm as} \ \tau \downarrow 0.
\end{eqnarray}

Therefore, gathering these asymptotics, we obtain
\begin{eqnarray}
Z[S^k K3](\tau,\frac{1}{2}) \rightarrow c(k)(-i \tau)^{0}
e^{\frac{\pi i k}{2 \tau}} \qquad {\rm as} \ \tau \downarrow 0.
 \quad \left( c(k) = (-1)^k {\tilde c}(k) \right) \label{eq:asymz}
\end{eqnarray}

The elliptic genus has the Fourier expansion of the form
\begin{eqnarray}
Z[S^k K3](\tau,\frac{1}{2})=\sum_{n=0}^{\infty} a_{n}q^n.
\label{eq:expgenus}
\end{eqnarray}
Again, due to the structure theorem, the coefficients satisfy $a_{n}
\leq a_{n+1}$. (See Appendix A for the explicit Fourier expansions of 
various functions.) Each coefficient $a_{n}$ represent the number of
the 1/2 BPS states of $h=n+k/4$ in the N=(4,4) $\sigma$-model.
This is the number of
the $1/8$ BPS states of the mass specified by
$h=n+k/4$ in the D1-D5 system \cite{Strominger-Vafa,Maldacena-Strominger}.
We can estimate the asymptotic form of $a_{n}$ by using the following
Tauberian theorem\cite{Kac1,Kac2}.

\noindent 
{\bf Theorem 3.} Let $f(\tau)$ be a function 
\begin{eqnarray*}
f(\tau) = q^{\lambda} \sum_{n=0}^{\infty} a_{n} q^n \qquad (q=e^{2 \pi
i \tau})
\end{eqnarray*}
which satisfies following conditions:
\begin{enumerate}
\renewcommand{\labelenumi}{\theenumi}
\renewcommand{\theenumi}{(\roman{enumi})}
\item $f(\tau)$ is a holomorphic function on ${\bf H}^{+}$.
\item $a_{n} \in {\bf R}$ and $a_{n} \leq a_{n+1}$ for all $n$.
\item There exist $c \in {\bf C}$, $d \in {\bf R}$ and $N \in 
      {\bf R}_{>0}$ such that 
      \begin{eqnarray*}
      f(\tau) \rightarrow c (-i \tau)^{-d} e^{\frac{2 \pi i N}{\tau}}
      \qquad {\rm as} \ \tau \downarrow 0.
      \end{eqnarray*}
\end{enumerate}
Then, the behavior of $a_{n}$ at large $n$ is
\begin{eqnarray*}
a_{n} \sim \frac{c}{\sqrt{2}}N^{-\frac{1}{2}(d-\frac{1}{2})}
n^{\frac{1}{2}(d-\frac{3}{2})} e^{2 \pi \sqrt{4 N n}}  \qquad {\rm as} \ 
n \rightarrow \infty, 
\end{eqnarray*}
where $a_{n} \sim b_{n}$ as $n \rightarrow \infty$ means $\lim_{n
\rightarrow \infty} b_{n}/a_{n} = 1$.  

Due to this theorem, the asymptotic form of $a_{n}$ can be read from
the estimation (\ref{eq:asymz})\footnote{We can
expect $c(k)$ is not large number due to the explicit
example of lower $k$. (See \cite{Kawai2}.)}
\begin{eqnarray}
a_{n} \sim \frac{c(k)}{\sqrt{2}} \left( 
\frac{k}{4} \right)^{\frac{1}{4}} n^{-\frac{3}{4}} e^{2 \pi \sqrt{k
n}}. \label{eq:bpsnum}
\end{eqnarray}  

\subsection{Counting the massive primary states}
We can also expand the elliptic genus by the characters of N=4 SCA
\begin{eqnarray}
&&\hspace{-2cm}Z[S^k K3](\tau,\frac{1}{2}) \nonumber \\ 
=&&\hspace{-0.5cm}\chi(S^k K3) \ {\rm ch}_{0}^{(R) k} 
(h=\frac{k}{4},l=0;\tau,0) + 
\sum_{l=\frac{1}{2}}^{\frac{k}{2}}F_{l}(\tau) \ {\rm ch}^{(R) k}
(h=\frac{k}{4},l;\tau,0).
\label{eq:charexp}
\end{eqnarray}
Each coefficient of $F_{l}(\tau) =
\sum_{n=0}^{\infty}a_{n}^{(l)} q^{n}$ counts the number of the 
massive representation ${\cal M}_{(h{\rm =}n+k/4,l)}^{k}$ 
in $Z[S^k K3](\tau,z)$
of the N=(4,4) $\sigma$-model. 
In particular $\sum_{l}a_{N}^{(l)}$ will be identified
with the number of the primary state (\ref{eq:extprimary}).
To obtain the asymptotic form $\sum_{l}a_{n}^{(l)}$, we may again
utilize the Tauberian theorem. For this purpose we need to know the
asymptotic behavior of $\sum_{l}F_{l}(\tau)$ as $\tau \downarrow 0$.

Since we have obtained the asymptotics of the elliptic genus
(\ref{eq:asymz}), the asymptotics of $\sum_{l}F_{l}(\tau)$ becomes
tractable if we can properly estimate the constituents of the
massless and massive characters. 
Let us remind that the character of 
massive representation ${\cal M}_{(h,l)}^{k}$ is
given by eq.(\ref{eq:charmass}).
The asymptotic behavior of
the character of $\widehat{{\rm SU}(2)}_{k}$ 
of the isospin $l$, eq.(\ref{eq:charaffine}), is given by\cite{Kac1,Kac2}
\begin{eqnarray}
\chi_{k}^{l}(\tau, 0) \rightarrow a(k,l) \exp \left( 
\frac{\pi i}{12 \tau} c_{k} \right)
\qquad {\rm as} \ \tau \downarrow 0,
\end{eqnarray}
where
\begin{eqnarray}
a(k,l) &=& \sqrt{\frac{2}{k+2}}\sin\left(\frac{(2 l + 1) \pi}{k+2}\right),
\nonumber \\
c_{k} &=& \frac{3 k}{k+2}.
\end{eqnarray}
Therefore, combining those of $\theta_{2}(\tau,0)$ and $\eta(\tau)$,
we obtain
\begin{eqnarray}
{\rm ch}^{(R) k}(h=\frac{k}{4},l;\tau,0) \rightarrow 
a(k{\rm -}1,l{\rm -}\frac{1}{2}) (-i \tau)^{\frac{1}{2}} \exp \left(
\frac{\pi i}{12 \tau} (3 + c_{k-1}) \right) \quad 
{\rm as} \ \tau \downarrow 0. \label{eq:asymchar}
\end{eqnarray}
 
As for the character of massless
representation ${\cal M}_{0 (l{\rm =}0)}^{k}$,
we can obtain the upper bound of the asymptotic
behavior by means of eq.(\ref{eq:relation}):
\begin{eqnarray}
\hspace{-1cm}{\rm ch}_{0}^{(R) k}
(h=\frac{k}{4},l=0;\tau,0)|_{\tau \downarrow 0}
&\leq& {\rm ch}^{(R) k}(h=\frac{k}{4},l=\frac{1}{2};\tau,0)|_{\tau
\downarrow 0} \nonumber \\
\hspace{-1cm}&=& a(k,l{\rm =}0)(-i \tau)^{\frac{1}{2}}\exp
\left( \frac{\pi i}{12 \tau} (3 + c_{k-1}) \right),
\end{eqnarray}
where $f(\tau)|_{\tau \downarrow 0}$ means the leading asymptotic of
$f(\tau)$ as $\tau \downarrow 0$.
From this estimation, the dominant contribution of the 
asymptotic behavior of $Z[S^k K3](\tau,1/2)$ turns out to come
from the part of the massive representations. 
We can neglect the contribution of the massless representations 
in the asymptotics.

Now we can obtain the asymptotic behavior of 
$\sum_{l}F_{l}(\tau)$.
Taking the limit $\tau \downarrow 0$ in eq.(\ref{eq:charexp}), 
\begin{eqnarray}
Z[S^k K3](\tau,\frac{1}{2}) \rightarrow 
\sum_{l=\frac{1}{2}}^{\frac{k}{2}} {\tilde F}_{l}(\tau)|_{\tau
\downarrow 0} \times (-i \tau)^{\frac{1}{2}}\exp \left( \frac{\pi i}{12 
\tau} (3+c_{k-1}) \right) \qquad {\rm as} \ \tau
\downarrow 0, \label{eq:asymf} 
\end{eqnarray}
where ${\tilde F}_{l}(\tau) = \sum_{n=0}^{\infty} {\tilde
a}_{n}^{(l)}q^{n} = a(k{\rm -1},l{\rm -1/2})F_{l}(\tau)$.
The asymptotic behavior of 
$\sum_{l}{\tilde F}_{l}(\tau)$
as $\tau \downarrow 0$ can be read by comparing (\ref{eq:asymf}) with
(\ref{eq:asymz}) 
\begin{eqnarray}
\sum_{l=\frac{1}{2}}^{\frac{k}{2}} {\tilde F}_{l}(\tau) \rightarrow
\ c(k)(-i \tau)^{- \frac{1}{2}} \exp \left( \frac{\pi i (6 k -
(3+c_{k-1}))}{12 \tau} \right) \quad {\rm as} \ \tau \downarrow 0.
\end{eqnarray}

$\sum_{l}{\tilde F}_{l}(\tau)$ has the
Fourier expansion of the form\footnote{These coefficients also satisfy 
$b_{n} \leq b_{n+1}$.} 
\begin{eqnarray}
\sum_{l=\frac{1}{2}}^{\frac{k}{2}} {\tilde F}_{l}(\tau) = 
\sum_{n=0}^{\infty} b_{n}q^n.
\qquad \left( b_{n} = \sum_{l=\frac{1}{2}}^{\frac{k}{2}} {\tilde
a}_{n}^{(l)} \right) \label{eq:expf}
\end{eqnarray} 
Due to the Tauberian theorem, the asymptotic form of
$b_{n}$ becomes:
\begin{eqnarray} 
b_{n} = \sum_{l=\frac{1}{2}}^{\frac{k}{2}} {\tilde a}_{n}^{(l)} \sim
\frac{c(k)}{\sqrt{2}}n^{- \frac{1}{2}} \exp \left( 2 \pi
\sqrt{(k-\frac{(3+c_{k-1})}{6}) n} \right) \qquad {\rm as} \ n \rightarrow
\infty. \label{eq:numpri}
\end{eqnarray}
Therefore, we conclude that the degeneracy of the massive
primary state of the dimension $h=n+k/4$ 
at $n\rightarrow\infty$ is given by eq.(\ref{eq:numpri}).

According to the argument in section 3.2, the degeneracy of the 
primary state (\ref{eq:prisigma}) corresponds to the degeneracy 
of the microscopic states of the extremal BTZ black hole 
${\rm BTZ}_{(N,N/l)}$. At the limit $N\gg k \gg 1$\footnote{The 
difference between $\sum_{l}{\tilde a}_{n}^{(l)}$ and
$\sum_{l}a_{n}^{(l)}$ is 
irrelevant in this semiclassical limit.}, that is, the semiclassical limit of
three-dimensional gravity, the degeneracy of the state 
(\ref{eq:extprimary}) becomes   
\begin{eqnarray} 
\sum_{l=\frac{1}{2}}^{\frac{k}{2}} {\tilde a}_{N}^{(l)} \sim
\frac{c(k)}{\sqrt{2}}N^{- \frac{1}{2}} \exp (2 \pi
\sqrt{(k-1) N}) \qquad {\rm as} \ N \rightarrow \infty, \label{eq:microent}
\end{eqnarray}
where $k=Q_{1}Q_{5}+1$.
The logarithm of eq.(\ref{eq:microent}) can be regarded as
the microscopic entropy of the extremal BTZ black hole with $Ml=J=N$.
It becomes
\begin{eqnarray}
S_{{\rm micro}} = 2 \pi \sqrt{Q_{1}Q_{5}N} + {\cal O}(\log N, \
\log c(k)).
\end{eqnarray} 
This completely agrees with the entropy formula
eq.(\ref{eq:extentropy}). 
This provides a justification of the
identification of the extremal BTZ black hole states with
the primary states (\ref{eq:extprimary}) of the N=(4,4) $\sigma$-model.

\section{Discussion} 

Until now, our study is limited to the case of the extremal BTZ black holes.
The non-extremal BTZ black holes can also appear as the near
horizon geometry of the non-BPS solitonic solutions in IIB supergravity.
According to the duality, the non-extremal BTZ black
holes may be also identified with the Virasoro primary states 
\begin{eqnarray}
   |h,l=0\rangle \otimes |{\tilde h},{\tilde l}=0\rangle  \ \quad {\rm with}
   \quad h, {\tilde h} > \frac{k}{4}, \label{eq:nonextpri}
\end{eqnarray}
in the corresponding N=(4,4) $\sigma$-model.

These states are in the tensor product of the massive representations both in
the left and right moving sectors.
So we must consider not the elliptic genus but the full
partition function of the N=(4,4) $\sigma$-model for the counting of
the degeneracy of the states.
It is known that the full partition function, which depends on the moduli
of $S^k K3$, has the contributions from the massive primary states of
$h={\bf Q}_{>0}+k/4$ (${\bf Q}$ : rational numbers)\cite{Eguchi4}. 
Therefore, the patition
function and the counterparts of $\sum_{l}F_{l}(\tau)$ can not have the
forms (\ref{eq:expgenus}) and (\ref{eq:expf}). So the Tauberian
theorem can not be applied to the counting of the primary state 
(\ref{eq:nonextpri}). 
We need the further investigations
for the well-defined counting of the microscopic states of 
the non-extremal BTZ black holes. 

Through this paper, we have discussed only the case of $M_{4}=K3$. 
The similar arguments in section 2.2 hold for        
the case of $M_{4}=T^4$. However the elliptic genus of the
corresponding N=(4,4) $\sigma$-model vanishes identically, since
this $\sigma$-model has the extra $U(1)^{4}$ symmetry other than
N=4 superconformal symmetry. So one cannot count the degeneracy
of the state (\ref{eq:extprimary}) by means of the elliptic genus.
In \cite{Maldacena3}, the counting of the 1/8 BPS states has been
argued by using another topological index called new supersymmetric
index. And they pointed out that the representations
of large N=4 superconformal algebra must be considered.
We may expect that the similar argument in this paper can be carried out
via the relation between this new supersymmetric
index and the characters of large N=4 superconformal algebra.

{\bf Acknowledgements} \\
The authors thank H. Umetsu and D. Tomino 
for their collaboration at the early stage of this work and 
also for their useful discussions and comments. 
The authors thank also T. Kawai for his useful comments.

\appendix
\section{Appendix}
Some formulas of elliptic theta functions and modular functions used
in the text are summarized\cite{Eldelyi,Kawai2}. 
  
Elliptic theta functions and Dedekind's $\eta$-function are 
defined by\footnote{$q=e^{2 \pi i \tau}$ and $y=e^{2 \pi i z}$.} 
\begin{eqnarray}
\theta_{1}(\tau,z) &=& i
\sum_{n=-\infty}^{\infty}(-1)^{n}q^{\frac{1}{2}(n-\frac{1}{2})^2}
y^{n-\frac{1}{2}}, \\
\theta_{2}(\tau,z) &=& \sum_{n=-\infty}^{\infty}
q^{\frac{1}{2}(n-\frac{1}{2})^2}y^{n-\frac{1}{2}} \ 
\left(= \theta_{1}(\tau,z+\frac{1}{2})\right),\\
\theta_{3}(\tau,z) &=& \sum_{n=-\infty}^{\infty}q^{\frac{1}{2}n^2}y^{n},
 \\
\theta_{4}(\tau,z) &=&
\sum_{n=-\infty}^{\infty}(-1)^{n}q^{\frac{1}{2}n^2}y^{n} \  
\left(=\theta_{4}(\tau,z+\frac{1}{2})\right), \\
\eta(\tau) &=& \sum_{n=-\infty}^{\infty}
(-1)^{n}q^{\frac{3}{2}(n+\frac{1}{6})^2} \ \left(=q^{\frac{1}{24}}
\prod_{n=1}^{\infty}(1-q^{n})\right).
\end{eqnarray}
Their modular transforms become 
\begin{eqnarray}
\theta_{1}(-\frac{1}{\tau},\frac{z}{\tau}) &=& -i 
(-i \tau)^{\frac{1}{2}} e^{\frac{\pi i z^2}{\tau}} \theta_{1}(\tau,z),
\quad \theta_{2}(-\frac{1}{\tau},\frac{z}{\tau}) =  
(-i \tau)^{\frac{1}{2}}e^{\frac{\pi i z^2}{\tau}}\theta_{4}(\tau,z),
\nonumber \\
\theta_{3}(-\frac{1}{\tau},\frac{z}{\tau}) &=&  
(-i \tau)^{\frac{1}{2}}e^{\frac{\pi i z^2}{\tau}}\theta_{3}(\tau,z),
\quad \theta_{4}(-\frac{1}{\tau},\frac{z}{\tau}) =  
(-i \tau)^{\frac{1}{2}}e^{\frac{\pi i z^2}{\tau}}\theta_{2}(\tau,z),
\nonumber \\
\eta(-\frac{1}{\tau}) &=& (-i \tau)^{\frac{1}{2}} \eta(\tau).
\end{eqnarray}

Weierstrass's $\wp$-function has the following form in terms of 
these functions.
\begin{eqnarray}
\wp(\tau,z) = - \frac{1}{3}\sum_{i=1}^{3}\left(
\frac{\theta_{i+1}(\tau,z)}{\theta_{i+1}(\tau,0)} \right)^{2} 
\frac{\eta(\tau)^6}{\theta_{1}(\tau,z)^2},
\end{eqnarray}
and, in particular, if we set $z=1/2$,
\begin{eqnarray}
\wp(\tau,\frac{1}{2}) = -\frac{1}{12} \left(\theta_{3}(\tau,0)^{4}+
\theta_{4}(\tau,0)^{4} \right).
\end{eqnarray}

The Eisenstein series of weight $k$ is defined by
\begin{eqnarray}
E_{k}(\tau) = 1 - \frac{2k}{B_{k}}\sum_{n=1}^{\infty}\sigma_{k-1}(n)
q^n \qquad (k \geq 4), 
\end{eqnarray}
where $B_{k}$ are the Bernoulli numbers and $\sigma_{k}(n) =
\sum_{d|n}d^k$.
These series satisfy following property under modular
transformation.
\begin{eqnarray}
E_{k}\left(\frac{a \tau + b}{c \tau + d}\right) = (c \tau + d)^{k}
E_{k}(\tau) \qquad (\pmatrix{a & b \cr c & d \cr} \in {\rm
SL}_{2}({\bf Z})).
\end{eqnarray}

\end{document}